\documentclass[twoside,twocolumn,9pt]{article}
\usepackage{extsizes}
\usepackage[super,sort&compress,comma]{natbib} 
\usepackage[version=3]{mhchem}
\usepackage[left=1.5cm, right=1.5cm, top=1.785cm, bottom=2.0cm]{geometry}
\usepackage{balance}
\usepackage{times,mathptmx}
\usepackage{sectsty}
\usepackage{graphicx} 
\usepackage{lastpage}
\usepackage[format=plain,justification=raggedright,singlelinecheck=false,font={stretch=1.125,small,sf},labelfont=bf,labelsep=space]{caption}
\usepackage{float}
\usepackage{fancyhdr}
\usepackage{fnpos}
\usepackage[english]{babel}
\usepackage{array}
\usepackage{droidsans}
\usepackage{charter}
\usepackage[T1]{fontenc}
\usepackage[usenames,dvipsnames]{xcolor}
\usepackage{setspace}
\usepackage[compact]{titlesec}
\usepackage{url}

\definecolor{cream}{RGB}{222,217,201}

\DeclareMathOperator{\Tr}{Tr}

\begin{document}

\pagestyle{fancy}
\thispagestyle{plain}
\fancypagestyle{plain}{

%%%HEADER%%%
%\fancyhead[C]{\includegraphics[width=18.5cm]{head_foot/header_bar}} %deleted
%\fancyhead[L]{\hspace{0cm}\vspace{1.5cm}\includegraphics[height=30pt]{head_foot/journal_name}} %deleted
%\fancyhead[R]{\hspace{0cm}\vspace{1.7cm}\includegraphics[height=55pt]{head_foot/RSC_LOGO_CMYK}} %deleted
\renewcommand{\headrulewidth}{0pt}
}
%%%END OF HEADER%%%

%%%PAGE SETUP - Please do not change any commands within this section%%%
\makeFNbottom
\makeatletter
\renewcommand\LARGE{\@setfontsize\LARGE{15pt}{17}}
\renewcommand\Large{\@setfontsize\Large{12pt}{14}}
\renewcommand\large{\@setfontsize\large{10pt}{12}}
\renewcommand\footnotesize{\@setfontsize\footnotesize{7pt}{10}}
\makeatother

\renewcommand{\thefootnote}{\fnsymbol{footnote}}
\renewcommand\footnoterule{\vspace*{1pt}% 
\color{cream}\hrule width 3.5in height 0.4pt \color{black}\vspace*{5pt}} 
\setcounter{secnumdepth}{5}

\makeatletter 
\renewcommand\@biblabel[1]{#1}            
\renewcommand\@makefntext[1]% 
{\noindent\makebox[0pt][r]{\@thefnmark\,}#1}
\makeatother 
\renewcommand{\figurename}{\small{Fig.}~}
\sectionfont{\sffamily\Large}
\subsectionfont{\normalsize}
\subsubsectionfont{\bf}
\setstretch{1.125} %In particular, please do not alter this line.
\setlength{\skip\footins}{0.8cm}
\setlength{\footnotesep}{0.25cm}
\setlength{\jot}{10pt}
\titlespacing*{\section}{0pt}{4pt}{4pt}
\titlespacing*{\subsection}{0pt}{15pt}{1pt}
%%%END OF PAGE SETUP%%%

%%%FOOTER%%%
\fancyfoot{}
\fancyfoot[LO,RE]{\vspace{-7.1pt}\includegraphics[height=9pt]{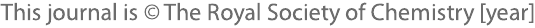}}
\fancyfoot[CO]{\vspace{-7.1pt}\hspace{13.2cm}\includegraphics{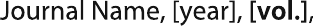}}
\fancyfoot[CE]{\vspace{-7.2pt}\hspace{-14.2cm}\includegraphics{head_foot/RF}}
\fancyfoot[RO]{\footnotesize{\sffamily{1--\pageref{LastPage} ~\textbar  \hspace{2pt}\thepage}}}
\fancyfoot[LE]{\footnotesize{\sffamily{\thepage~\textbar\hspace{3.45cm} 1--\pageref{LastPage}}}}
\fancyhead{}
\renewcommand{\headrulewidth}{0pt} 
\renewcommand{\footrulewidth}{0pt}
\setlength{\arrayrulewidth}{1pt}
\setlength{\columnsep}{6.5mm}
\setlength\bibsep{1pt}
%%%END OF FOOTER%%%

%%%FIGURE SETUP - please do not change any commands within this section%%%
\makeatletter 
\newlength{\figrulesep} 
\setlength{\figrulesep}{0.5\textfloatsep} 

\newcommand{\topfigrule}{\vspace*{-1pt}% 
\noindent{\color{cream}\rule[-\figrulesep]{\columnwidth}{1.5pt}} }

\newcommand{\botfigrule}{\vspace*{-2pt}% 
\noindent{\color{cream}\rule[\figrulesep]{\columnwidth}{1.5pt}} }

\newcommand{\dblfigrule}{\vspace*{-1pt}% 
\noindent{\color{cream}\rule[-\figrulesep]{\textwidth}{1.5pt}} }

\makeatother
%%%END OF FIGURE SETUP%%%

%%%TITLE, AUTHORS AND ABSTRACT%%%
\twocolumn[
  \begin{@twocolumnfalse}
{\includegraphics[height=30pt]{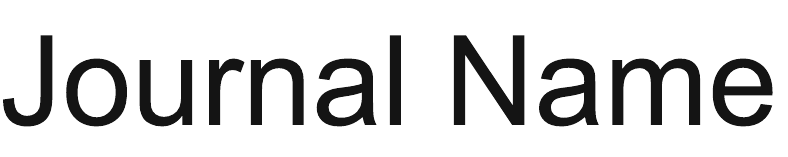}\hfill%added
 \raisebox{0pt}[0pt][0pt]{\includegraphics[height=55pt]{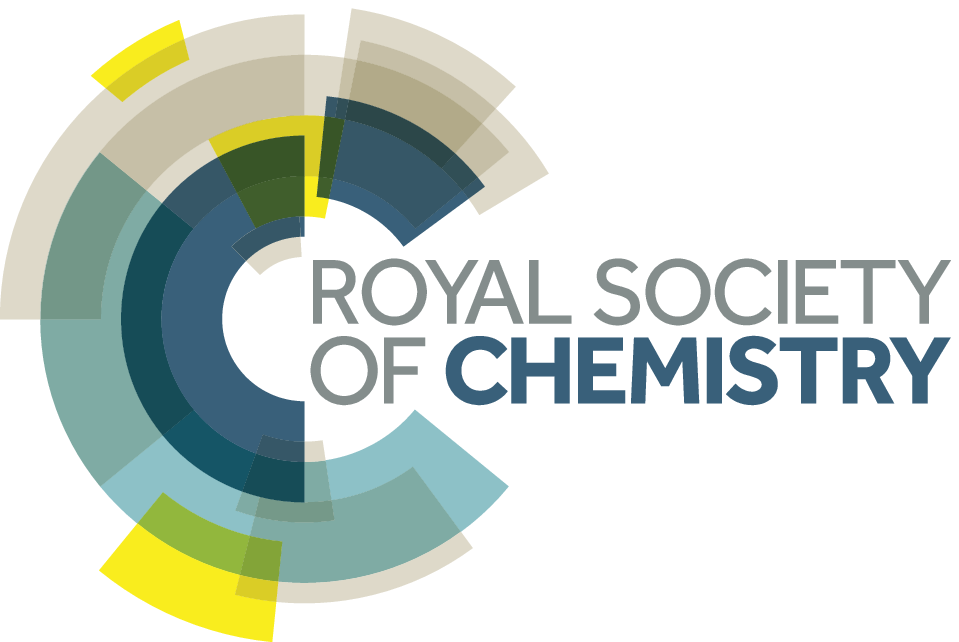}}%added
 \\[1ex]%added
 \includegraphics[width=18.5cm]{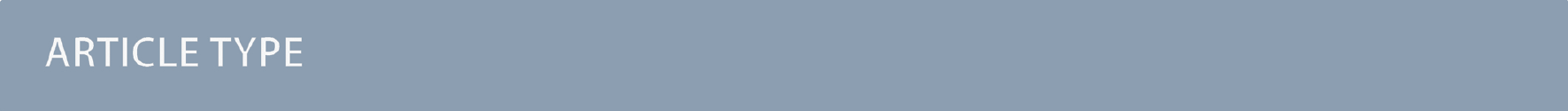}}\par%added
\vspace{1em}%added
\vspace{3cm}%deleted
\sffamily
\begin{tabular}{m{4.5cm} p{13.5cm} }

\includegraphics{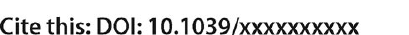} & \noindent\LARGE{\textbf{Explicit Demonstration of Geometric Frustration in Chiral Liquid Crystals}} \\%Article title goes here instead of the text "This is the title"
\vspace{0.3cm} & \vspace{0.3cm} \\

 & \noindent\large{Cheng Long\textit{$^{a}$} and Jonathan V. Selinger$^{\ast}$\textit{$^{a}$}} \\%Author names go here instead of "Full name", etc.

\includegraphics{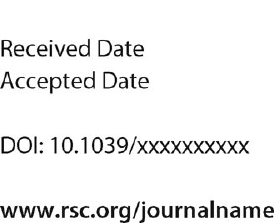} & \noindent\normalsize{Many solid materials and liquid crystals exhibit geometric frustration, meaning that they have an ideal local structure that cannot fill up space. For that reason, the global phase must be a compromise between the ideal local structure and geometric constraints. As an explicit example of geometric frustration, we consider a chiral liquid crystal confined in a long cylinder with free boundaries. When the radius of the tube is sufficiently small, the director field forms a double-twist configuration, which is the ideal local structure. However, when the radius becomes larger (compared with the natural twist of the liquid crystal), the double-twist structure cannot fill space, and hence the director field must transform into some other chiral structure that can fill space. This space-filling structure may be either (1) a cholesteric phase with single twist, or (2) a set of double-twist regions separated by a disclination, which can be regarded as the beginning of a blue phase. We investigate these structures using theory and simulations, and show how the relative free energies depend on the system size, the natural twist, and the disclination energy. As another example, we also study a cholesteric liquid crystal confined between two infinite parallel plates with free boundaries.} \\%The abstract goes here instead of the text "The abstract should be..."

\end{tabular}

 \end{@twocolumnfalse} \vspace{0.6cm}

  ]
%%%END OF TITLE, AUTHORS AND ABSTRACT%%%

%%%FONT SETUP - please do not change any commands within this section
\renewcommand*\rmdefault{bch}\normalfont\upshape
\rmfamily
\section*{}
\vspace{-1cm}

%%%FOOTNOTES%%%

\footnotetext{\textit{$^{a}$~Department of Physics, Advanced Materials and Liquid Crystal Institute, Kent State University, Kent, OH 44242, USA; E-mail: jselinge@kent.edu}}

%Please use \dag to cite the ESI in the main text of the article.
%If you article does not have ESI please remove the the \dag symbol from the title and the footnotetext below.
%\footnotetext{\dag~Electronic Supplementary Information (ESI) available: [details of any supplementary information available should be included here]. See DOI: 10.1039/b000000x/}
%additional addresses can be cited as above using the lower-case letters, c, d, e... If all authors are from the same address, no letter is required

%\footnotetext{\ddag~Additional footnotes to the title and authors can be included \emph{e.g.}\ `Present address:' or `These authors contributed equally to this work' as above using the symbols: \ddag, \textsection, and \P. Please place the appropriate symbol next to the author's name and include a \texttt{\textbackslash footnotetext} entry in the the correct place in the list.}

%%%END OF FOOTNOTES%%%

%%%MAIN TEXT%%%%

\section{Introduction}

In condensed matter physics, there are many situations where spatial geometry is not compatible with optimum local interactions.  In these cases, materials cannot fill up space with the ideal local structure, and must find complex ways to reconcile the incompatibility \cite{sadoc_mosseri_1999,ramirez2003geometric,han2008geometric,hirata2013geometric,selinger2022director}. A well-known example of geometric frustration is an Ising antiferromagnet on a triangular lattice. Once the first two spins on a triangle align opposite to each other, the third spin is frustrated because it cannot simultaneously minimize its interactions with the other two. This frustration leads to a highly degenerate ground state. Such geometric frustration in the spin system has many significant physical consequences, such as residual entropy in water ice and spin ice \cite{harris1997geometrical,ramirez2003geometric}. Other typical examples of geometric frustration can be found in close-packing problems in free space or confined geometries \cite{grason2015colloquium,hagan2021equilibrium}. As a result, the concept of geometric frustration is of crucial importance in understanding many complex natural phenomena. In some cases, geometric frustration can also be utilized to engineer structures in condensed matter systems \cite{araki2013defect}. For example, in a nematic liquid crystal, boundary conditions can be designed intentionally to introduce topological defects in the bulk.

In a recent article \cite{selinger2022director}, our group argued that chiral liquid crystals are geometrically frustrated. Our argument is based on a reformulated version of the Oseen-Frank elastic theory \cite{selinger2018interpretation}, which decomposes the free energy density into four fundamental elastic modes---the well-known splay, twist, and bend modes, and the less-well-known biaxial splay ($\boldsymbol{\Delta}$) mode. In this reformulated theory, the saddle-splay free energy is naturally included as a combination of the bulk elastic modes. The theory shows that a chiral liquid crystal has a natural tendency to form pure twist, which is double twist. However, it is impossible to fill up 3D Euclidean space with pure double twist \cite{virga2019uniform}. One way for a chiral liquid crystal to reconcile this incompatibility is to form the cholesteric phase, which has a combination of the favorable twist and the unfavorable $\boldsymbol{\Delta}$ mode. Another way is to form a blue phase, which has regions of approximately pure twist separated by disclinations. Hence, both the cholesteric phase and blue phases can be regarded as frustrated structures.

This view of the cholesteric phase as a frustrated structure may be surprising to some readers. Researchers often neglect saddle-splay and the $\boldsymbol{\Delta}$ mode, and consider only splay, twist, and bend. In this common way of thinking, the cholesteric phase appears to be an ideal unfrustrated structure, because it has uniform nonzero twist, with zero splay and bend.

The purpose of this article is to demonstrate explicitly that the common way of thinking is incorrect, and the cholesteric phase really is frustrated. For this demonstration, we consider a chiral liquid crystal in finite geometries with free boundary conditions. As a general rule, one can identify the ideal local structure of a material by looking near a free boundary. Any material feels strong packing constraints in the interior, but it has much more freedom to achieve its ideal local structure at the boundary. Here, we show that a chiral liquid crystal has double twist, not cholesteric single twist, near a free boundary. For a small system size, double twist extends throughout the system. For a large system size, double twist exists in a boundary layer, while cholesteric single twist may form in the interior.

In Sec.~II, we review the reformulated Oseen-Frank theory and show that the optimum local deformation is double twist, not cholesteric single twist. We then consider two specific finite geometries with free boundary conditions. In Sec.~III, we model a chiral liquid crystal in a cylinder with free boundary conditions, using a combination of approximate analytic calculations, director simulations, and nematic order tensor simulations. When the cylinder radius is small, compared with the natural twist of the liquid crystal, the ground state has double twist.  When the radius increases, geometric frustration causes a transition to a different configuration---either single twist (as in the cholesteric phase) or double twist with disclinations (as in a blue phase). In Sec.~IV, we study a chiral liquid crystal between two parallel surfaces with free boundary conditions, using director simulations. In this case, geometric frustration induces distortions close to the free surfaces, leading to surface states composed of regularly arranged double-twist regions.

\section{Optimum Local Deformation}

We consider a chiral nematic liquid crystal described by the director field $\hat{\boldsymbol{n}}(\boldsymbol{r})$. As discussed in previous papers \cite{selinger2018interpretation,selinger2022director}, the reformulated Oseen-Frank free energy density takes the form
\begin{eqnarray}\label{Selinger_free_energy}
  \nonumber f &=& \frac{1}{2}\left(K_{11}-K_{24}\right) S^2 + \frac{1}{2}\left(K_{22}-K_{24}\right) T^2 + \frac{1}{2} K_{33} |\boldsymbol{B}|^2 \\
  & & + K_{24} \mathrm{Tr}\left( \boldsymbol{\Delta}^2 \right) - K_{22} q_0 T .
\end{eqnarray}
In this expression, the first three terms represent the free energy cost of splay $S=\boldsymbol{\nabla} \cdot \hat{\boldsymbol{n}}$, twist $T=\hat{\boldsymbol{n}} \cdot \left(\boldsymbol{\nabla} \times \hat{\boldsymbol{n}}\right)$, and bend $\boldsymbol{B}=\hat{\boldsymbol{n}} \times \left(\boldsymbol{\nabla} \times \hat{\boldsymbol{n}}\right)$ deformations. The fourth term represents the free energy cost of the biaxial splay $\boldsymbol{\Delta}$, which is the traceless, symmetric tensor 
\begin{eqnarray}\label{biaxial_splay_mode}
  \nonumber \Delta_{ij} &=& \frac{1}{2}(\partial_i n_j + \partial_j n_i - n_i n_k \partial_k n_j - n_j n_k \partial_k n_i \\
  & & - \delta_{ij}\partial_k n_k + n_i n_j \partial_k n_k) .
\end{eqnarray}
The last term is an extra contribution to the free energy, linear in the twist, which is permitted by symmetry in a chiral liquid crystal. Here, we write its coefficient as $K_{22}q_0$, where $q_0$ is an inverse length. We will see below that $q_0$ is related to the natural pitch of the chiral liquid crystal.

In this section, we want to minimize the \emph{local} free energy density at a single point in the liquid crystal. We choose coordinates so that this point is the origin, and the director at that point is $\hat{\boldsymbol{n}}=\hat{\boldsymbol{z}}$. One possible configuration of the director field is a cholesteric phase with single twist. If we choose coordinates so that the helical axis is along $\hat{\boldsymbol{x}}$, the director field can be written as
\begin{eqnarray}\label{cholesteric_helical_structure}
  \hat{\boldsymbol{n}}_{ST} = \hat{\boldsymbol{y}}\sin q x + \hat{\boldsymbol{z}}\cos q x,
\end{eqnarray}
with the subscript $ST$ for single twist. This director field has the deformation modes
\begin{eqnarray}\label{DeltaMode_cholesteric_helical_structure}
  && S=0, \quad T=q, \quad \boldsymbol{B}=0,\\
  \nonumber && \boldsymbol{\Delta} = 
  \begin{pmatrix}
  0 & \frac{1}{2}q\cos{q x} & -\frac{1}{2}q\sin{q x} \\
  \frac{1}{2}q\cos{q x} & 0 & 0 \\
  -\frac{1}{2}q\sin{q x} & 0 & 0
  \end{pmatrix}.
\end{eqnarray}
By inserting these deformation modes into Eq.~(\ref{Selinger_free_energy}), we find the free energy density $f_{ST}=\frac{1}{2}K_{22}q^2-K_{22} q_0 q$.  Minimizing that expression over the parameter $q$ then gives $q=q_0$ and
\begin{eqnarray}\label{fST}
  f_{ST}=-\frac{K_{22}q_0^2}{2}.
\end{eqnarray}

As an alternative, the director field could form a configuration with double twist. In cylindrical coordinates, that configuration can be written as
\begin{eqnarray}\label{double_twist_structure}
  \hat{\boldsymbol{n}}_{DT} = \hat{\boldsymbol{\phi}}\sin\theta(\rho) + \hat{\boldsymbol{z}}\cos\theta(\rho),
\end{eqnarray}
with the subscript $DT$ for double twist. The function $\theta(\rho)$ describes how the director depends on the radial coordinate away from the $z$-axis. At the origin, the director is aligned with the $z$-axis, and hence $\theta(0) = 0$. Near the origin, $\theta(\rho)$ can be approximated by the lowest-order term in a power series, $\theta(\rho) \approx q \rho$. With this assumption, the four elastic modes at the origin are $T=2q$, $S=0$, $\boldsymbol{B}=0$, and $\boldsymbol{\Delta}=0$. Hence, the free energy density at the origin is $f_{DT}=2(K_{22}-K_{24}) q^2 - 2 K_{22} q_0 q$. Minimizing this free energy over the parameter $q$ gives $q=K_{22} q_0/[2(K_{22}-K_{24})]$ and 
\begin{eqnarray}\label{fDT}
  f_{DT}=-\frac{K_{22}^2 q_0^2}{2(K_{22}-K_{24})}.
\end{eqnarray}

Now we can compare the local free energy densities $f_{ST}$ and $f_{DT}$. From Eqs.~(\ref{fST}) and~(\ref{fDT}), we can see that the double-twist configuration has a lower free energy density when ${K_{24}>0}$.  By comparison, the single-twist configuration has a lower free energy density when ${K_{24}<0}$, and the configurations are degenerate when ${K_{24}=0}$.

In most ordinary liquid crystals, all four quadratic terms in the free energy density of Eq.~(\ref{Selinger_free_energy}) are positive, and hence the elastic coefficients satisfy the conditions $(K_{11}-K_{24})>0$, $(K_{22}-K_{24})>0$, $K_{33}>0$, and $K_{24}>0$. These conditions are called the Ericksen inequalities \cite{Ericksen1966,selinger2018interpretation}. Some recent studies have shown that the inequality $(K_{22}-K_{24})>0$ can be violated in lyotropic chromonic liquid crystals \cite{davidson2015chiral,nayani2015spontaneous,long2022violation}. However, to our knowledge, there are no reports of liquid crystals violating the Ericksen inequality $K_{24}>0$. Indeed, any violation of that inequality would lead to a peculiar liquid crystal with a spontaneous $\boldsymbol{\Delta}$ deformation. We must conclude that $K_{24}>0$ is the general case, and thus the local free energy density is lower for double twist than for single twist.

Although double twist is preferred over single twist locally, at a single point, we do not yet know what will happen over longer length scales. Mathematical studies \cite{virga2019uniform,pollard2021intrinsic,da2021moving} have proved that pure double twist cannot fill up 3D Euclidean space; i.e.\ it is not compatible with 3D Euclidean geometry. Instead, over longer length scales, double twist must be combined with the other deformation modes. To find the effects of these geometric constraints, we must consider a finite system with a specific size and shape, and we must minimize the integrated free energy rather than the free energy density.

In the following sections, we investigate two finite systems---a long cylinder and a large slab. In order to eliminate the possibility of introducing deformations from anchoring conditions \cite{oswald2005nematic,araki2013defect}, we only consider free boundary conditions.

\section{Chiral Liquid Crystal in a Cylinder}

In this section, we investigate a chiral liquid crystal confined in a long cylinder of radius $R$ with its long axis along $\hat{\boldsymbol{z}}$. The cylinder has free boundary conditions, so that the director on the boundary does not have any preferred orientation. To simplify our model, we only consider the director field as a function of $x$ and $y$, assuming that it is independent of $z$.

\subsection{Approximate Analytic Calculation}

We begin with an approximate analytic calculation to compare the total free energies of single-twist and double-twist configurations inside the finite cylinder.

For single twist, we assume the director field of Eq.~(\ref{cholesteric_helical_structure}).  This director field fills up space, with the free energy density given by Eq.~(\ref{fST}), independent of position.  Hence, the total free energy (per length in the $z$ direction) is just
\begin{eqnarray}\label{Ftotal_single_twist}
  F_{\mathrm{ST}} = -\frac{K_{22}q_0^2}{2} \pi R^2 = -\frac{\pi K_{22}\bar{q}_0^2}{2}.
\end{eqnarray}
Here, $\bar{q}_0=q_0 R$ is a dimensionless parameter representing the natural twist $q_0$ of the liquid crystal, scaled by the cylinder radius $R$.

For double twist, we assume the director field of Eq.~(\ref{double_twist_structure}), in cylindrical coordinates, with an unknown function $\theta(\rho)$.  The total free energy (per length in the $z$ direction) then becomes the integral
\begin{align}
  F_{\mathrm{DT}} &= \int_0^R 2\pi\rho d\rho\Biggl[
  \frac{K_{22}}{2}\left(\theta'+\frac{\sin2\theta}{2\rho}\right)^2
  +\frac{K_{33}\sin^4\theta}{2\rho^2}\nonumber\\
  &\quad-\frac{K_{24}\theta'\sin2\theta}{\rho}
  -K_{22}q_0\left(\theta'+\frac{\sin2\theta}{2\rho}\right)\Biggr].
  \label{fDTintegral}
\end{align}
To minimize the free energy, we derive the Euler-Lagrange equation
\begin{equation}
  \theta''+\frac{\theta'}{\rho}-\frac{\sin4\theta}{4\rho^2}
  -\frac{2K_{33}\cos\theta\sin^3\theta}{K_{22}\rho^2}-\frac{2q_0\sin^2\theta}{\rho}=0.
\end{equation}
At $\rho=0$, the boundary condition is $\hat{\boldsymbol{n}}=\hat{\boldsymbol{z}}$, so that
\begin{equation}
  \theta(0)=0.
\end{equation}
At $\rho=R$, we have a free boundary, which implies that $\partial f/\partial\theta'=0$, and hence
\begin{equation}
  \theta'(R)+\frac{\sin2\theta(R)}{2R}-\frac{K_{24}\sin2\theta(R)}{K_{22}R}-q_0=0.
\end{equation}
We solve this system of equations as a power series in the natural twist $q_0$, which gives
\begin{align}
\theta(\rho)&=\frac{q_0\rho K_{22}}{2(K_{22}-K_{24})}
+\frac{q_0^3\rho^3 K_{22}^2 (2K_{22}-6K_{24}+3K_{33})}{96(K_{22}-K_{24})^3}\nonumber\\
&+\frac{q_0^3 R^2 \rho K_{22}^2 (2K_{22}K_{24}-2K_{24}^2-2K_{22}K_{33}+K_{24}K_{33})}{32(K_{22}-K_{24})^4}\nonumber\\
&+O(q_0^5).
\label{theta_of_rhobar_solution}
\end{align}
We can see that $\theta(\rho)$ is an odd function of $q_0$, because switching the sign of the natural twist reverses the handedness of the resulting configuration. This solution implies a director field with twist of order $q_0$, bend of order $q_0^2$, $\boldsymbol{\Delta}$ mode of order $q_0^3$, and zero splay.

As an aside, the factors of $(K_{22}-K_{24})$ in the denominators of Eq.~(\ref{theta_of_rhobar_solution}) show that the series expansion breaks down if $K_{24}\to K_{22}$. In this limit, the Ericksen inequality $K_{22}-K_{24}>0$ is violated, and the liquid crystal is at a critical point for the formation of spontaneous double twist, as discussed in Ref.~\cite{long2022violation}. At the critical point, it has a divergent susceptibility to $q_0$, which might be regarded as a applied chiral field. In that limit, we find $\theta(\rho)\propto (q_0)^{1/3}$. We will not consider that special case further in this article.

To find the total free energy of the double-twist configuration (per length in the $z$ direction), we put the series expansion for $\theta(\rho)$ back into Eq.~(\ref{fDTintegral}).  This calculation gives
\begin{equation}\label{Ftotal_double_twist}
  F_{\mathrm{DT}} = -\frac{\pi K_{22}^2\bar{q}_0^2}{2(K_{22}-K_{24})}
  + \frac{\pi K_{22}^4 K_{33}\bar{q}_0^4}{64(K_{22}-K_{24})^4}+O(\bar{q}_0^6),
\end{equation}
where again $\bar{q}_0=q_0 R$. Now we can compare the total free energies of the single-twist configuration in Eq.~(\ref{Ftotal_single_twist}) and the double-twist configuration in Eq.~(\ref{Ftotal_double_twist}). When the parameter $\bar{q}$ is small, double twist has a lower free energy than single twist, as expected from the previous section. However, when $\bar{q}$ increases, the positive quartic term raises the free energy for double twist. If we neglect higher-order terms in the power series, we can estimate that the free energies become equal at
\begin{eqnarray}\label{transition_point}
  \bar{q}_0 = 2^{5/2}\left(\frac{K_{22}-K_{24}}{K_{22}}\right)^{3/2}\left(\frac{K_{24}}{K_{33}}\right)^{1/2}.
\end{eqnarray}
Beyond this transition point, single twist has a lower free energy than double twist. As a result, our free energy comparison predicts a transition between double twist and single twist at a particular $\bar{q}_0$, which can be induced by either changing the natural twist $q_0$ or changing the cylinder radius $R$.

In the case of equal Frank elastic constants, $K_{11}=K_{22}=K_{33}=2K_{24}=K$, our prediction for the transition point is just $\bar{q}_0=\sqrt{2}$.  If $K_{24}$ decreases toward 0, the double-twist regime becomes smaller and the single-twist regime becomes larger, so that it becomes easier for the liquid crystal to form a cholesteric phase with single twist. This trend is reasonable because the $K_{24}$ gives the energy cost for the unfavorable $\boldsymbol{\Delta}$ deformation in the cholesteric phase. By comparison, if $K_{33}$ decreases toward 0, the double-twist regime becomes larger and the single-twist regime becomes smaller. That trend is reasonable because $K_{33}$ gives the energy cost for the unfavorable bend deformation, which is present in the double-twist structure away from the central axis at $\rho=0$.

\subsection{Director Field Simulations}

We would like to go beyond the comparison of single-twist and double-twist configurations, in order to determine whether the liquid crystal can cross over between these limits by forming intermediate configurations.  For this reason, we perform simulations of the director field inside a cylinder.

In director field simulations, researchers often parameterize the three components of $\hat{\boldsymbol{n}}$ in terms of the angles $\theta$ and $\phi$ in conventional spherical coordinates.  Here, that parameterization is not convenient, because the coordinate system is singular whenever $\hat{\boldsymbol{n}}$ is along the $z$-axis, which occurs right in the center of the double-twist configuration.  As an alternative, we use a version of spherical coordinates based on the $x$-axis.  Hence, we write 
\begin{eqnarray}\label{director_field_n}
  \hat{\boldsymbol{n}}(x,y) = \left(\sin{u}, \cos{u}\sin{v}, \cos{u}\cos{v}\right),
\end{eqnarray}
in terms of angles $u(x,y)$ and $v(x,y)$.  We assume equal Frank elastic constants, $K_{11}=K_{22}=K_{33}=2K_{24}=K$, so that the free energy (per length in the $z$ direction) simplifies to
\begin{align}\label{ftotal_director_field_simulations}
  F &= K\int dx dy\biggl[\frac{1}{2}\left(\partial_i n_j\right) \left(\partial_i n_j\right)
  -q_0\epsilon_{ijk}n_i\partial_j n_k\biggr]\\
  &= K\int dx dy\biggl[\frac{1}{2}|\boldsymbol{\nabla}u|^2+\frac{1}{2}(\cos^2 u)|\boldsymbol{\nabla}v|^2
  +q_0(\cos v)\partial_y u
  \nonumber\\
  &\qquad\qquad
  -q_0(\cos^2 u)\partial_x v
  + \frac{1}{2}q_0(\sin 2u)(\sin v)\partial_y v\biggr].\nonumber
\end{align}
To simulate time-dependent relaxation, we define the dissipation function as
\begin{equation}\label{dissipation_function_for_director_field}
 D = \frac{1}{2} \gamma_1 \int dx dy  |\dot{\boldsymbol{n}}|^2
 = \frac{1}{2} \gamma_1 \int dx dy \left[\dot{u}^2 +(\cos^2 u)\dot{v}^2\right],
\end{equation}
with rotational viscosity $\gamma_1$.  We then solve the overdamped equations of motion
\begin{eqnarray}\label{equation_of_motion_director_field}
  -\frac{\delta F}{\delta u}-\frac{\delta D}{\delta\dot{u}}=0,\quad
  -\frac{\delta F}{\delta v}-\frac{\delta D}{\delta\dot{v}}=0,
\end{eqnarray}
by finite-element modeling using the software package COMSOL, running foward in time until the director field reaches an equilibrium state.  We choose units so that $K=\gamma_1=1$ and radius $R=10$, and vary the natural twist $q_0$.

Because the product $\bar{q}_0=q_0 R$ is the only dimensionless parameter in the problem, it must control the equilibrium state of the director field.  To determine how the equilibrium state evolves as a function of this parameter, we begin with a uniform nematic configuration aligned with the $z$-axis at $q_0 = 0$, then add a small increment to $q_0$ and let the director field relax. After a steady state is reached, another increment is added to $q_0$, and this process is repeated until $q_0$ is sufficiently large. To avoid trapping the director field in a metastable state, we also do simulations starting from the final state at the largest value of $q_0$, and then reducing $q_0$ by a series of decrements. We identify the ground state by comparing the total free energies of the different configurations at the same $q_0$. 

\begin{figure*}
\centering
\includegraphics[width=0.92\textwidth]{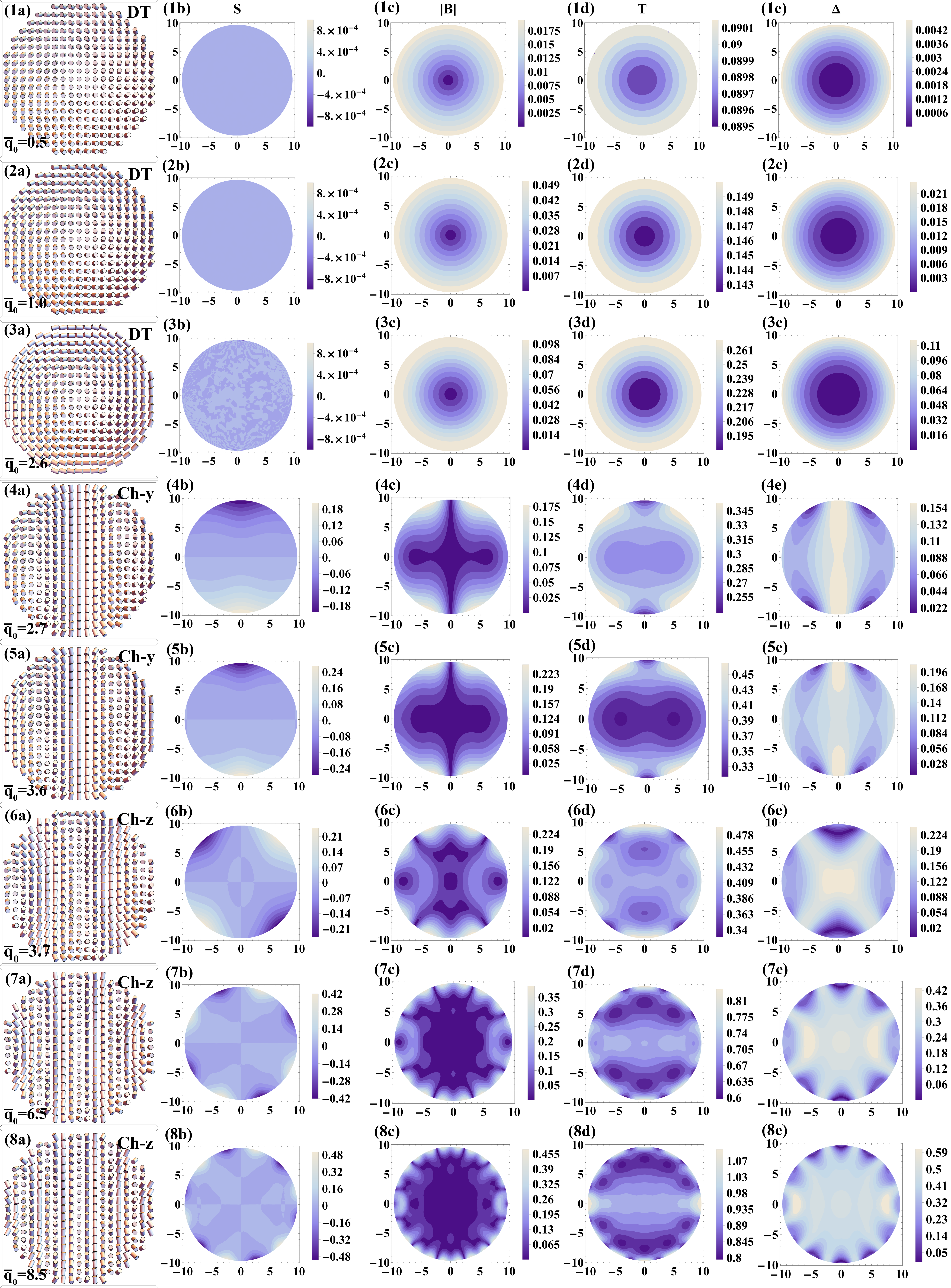}
\caption{Ground states of the simulated director field for different values of $\bar{q}_0=q_0 R$.  Rows 1--3 are the double-twist (DT) configuration, rows 4--5 are cholesteric-y (Ch-y), and rows 6-8 are cholesteric-z (Ch-z).  Column a shows visualizations of the director field, with cylinders representing the apolar director.  Columns b--e show contour plots of the splay, bend, twist, and $\boldsymbol{\Delta}$ modes, respectively.  (Column c represents the magnitude $|\boldsymbol{B}|$, and column e represents the positive eigenvalue of $\boldsymbol{\Delta}$.)}
\end{figure*}

The ground state configurations at different values of $\bar{q}_0$ are shown in Fig.~1. When $\bar{q}_0 = 0.5$, the chiral liquid crystal is most stable in a double-twist (DT) configuration (Fig.~1, row~1). By calculating the four elastic modes in the director field, we find that there is a large twist throughout the cylinder.  The bend and $\boldsymbol{\Delta}$ modes are zero at the center of the cylinder, but near the edge they grow to be nonzero (although much smaller than the twist).  The splay is zero everywhere.  All four elastic modes show full rotational symmetry about the $z$-axis. When $\bar{q}_0$ increases to $1.0$ or $2.6$ (Fig.~1, rows~2--3), the ground state remains a double-twist configuration. The observed twist increases because of the increased natural twist, and the bend and $\boldsymbol{\Delta}$ modes also grow substantially.  These features of the simulation results for small $\bar{q}_0$ are consistent with the predictions for the DT state from the approximate analytic calculation, except that DT persists up to a somewhat higher value of $\bar{q}_0$ than expected from that calculation.

When $\bar{q}_0$ increases from $2.6$ to $2.7$, a symmetry-breaking transition occurs, and the ground state transforms into a cholesteric-like configuration (Fig.~1, row~4). At this transition, the full rotational symmetry is lost.  In the bulk of the cylinder, away from the surface, the director configuration is similar to a cholesteric helical structure with its pitch axis along $\hat{\boldsymbol{x}}$. Because the director orientation at the center is along $\hat{\boldsymbol{y}}$, we call this structure cholesteric-y (Ch-y). From the contour plots of the four elastic modes, away from the surface, we see that the $\boldsymbol{\Delta}$ mode is about half of the twist, while the splay and bend modes are almost zero. These results agree with the predictions for cholesteric single twist. Close to the surface, the Ch-y configuration has a mixture of all four elastic modes. This configuration persists as $\bar{q}_0$ increases up to $3.6$  (Fig.~1, row~5).

When $\bar{q}_0$ increases from $3.6$ to $3.7$, another transition occurs, and the ground state becomes a cholesteric-z (Ch\nobreakdash-z) configuration (Fig.~1, row~6). In the bulk of the cylinder, the Ch-z director field is also a cholesteric helical structure similar to Ch-y, except that the director at the center is along $\hat{\boldsymbol{z}}$. Close to the surface, Ch-z shows a more complex behavior than the structure in the bulk. The free boundary condition allows formation of double twist, so that the director field introduces more double twist at the surface while being constrained by the single twist in the bulk. As $\bar{q}_0$ further increases to $6.5$ and $8.5$ (Fig.~1, rows~7--8), more cholesteric pitches are pushed into the cylinder, and the entire Ch-z configuration becomes more like a standard cholesteric helical structure.

To further understand the structural transitions inside the cylinder, we plot the free energies of the simulated DT, Ch-y, and Ch-z states as functions of $\bar{q}_0$ in Fig.~2(a). To emphasize the differences in free energies, we subtract off a common baseline, which is the free energy of the standard cholesteric single-twist configuration $F_{\mathrm{ST}}$ from Eq.~(\ref{Ftotal_single_twist}). This figure shows that the free energies of the configurations cross at two specific values of $\bar{q}_0$.  The first derivative of the total free energy changes discontinuously at the two transition points, which indicates that those two structural transitions are similar to first-order phase transitions. The discontinuity in slope is large at the transition from DT to CH-y, but much smaller at the transition from Ch-y to Ch-z, so that the latter could be a weak first-order transition.

\begin{figure}
\centering
\includegraphics[width=0.45\textwidth]{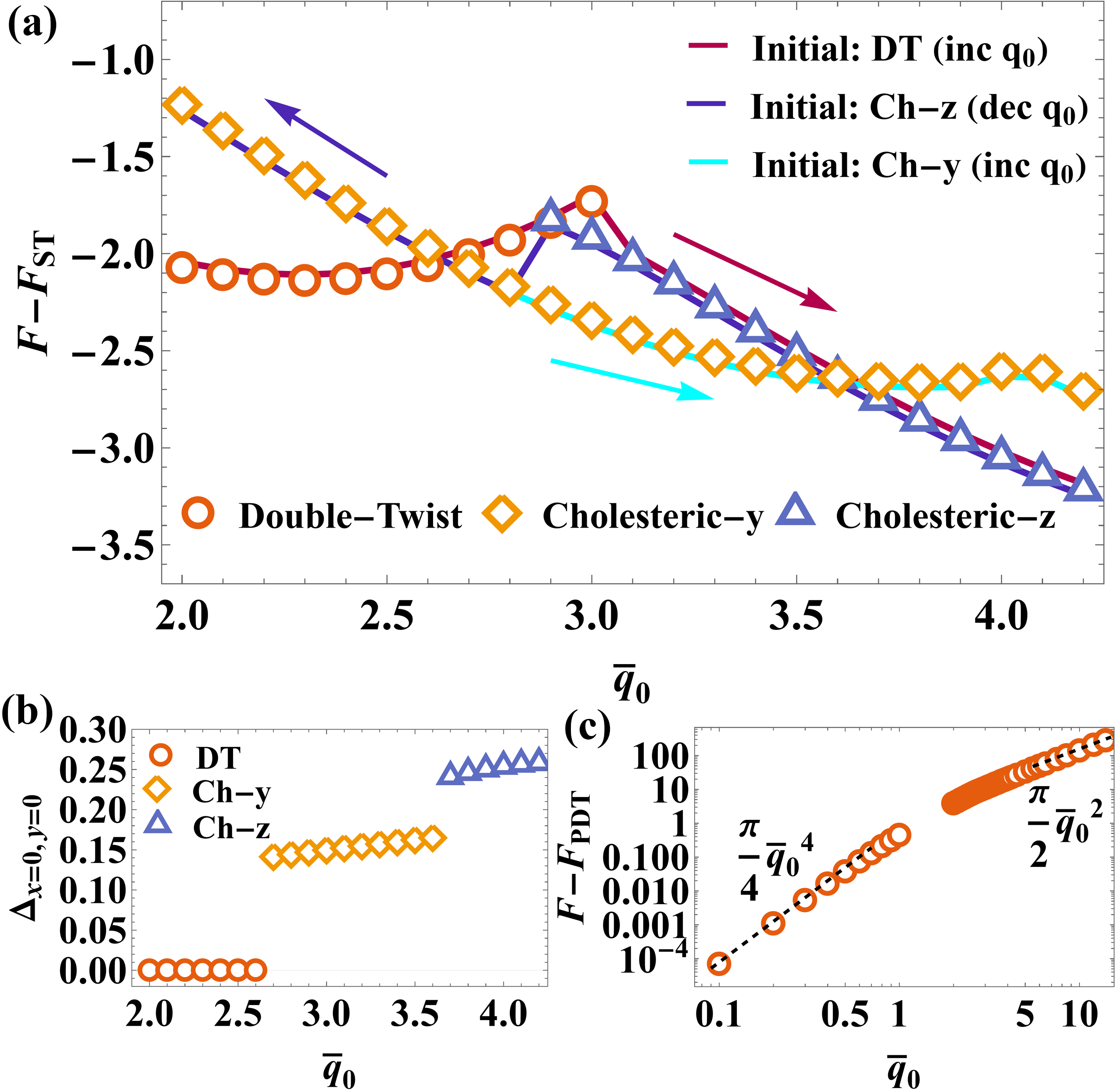}
\caption{(a)~Comparison of the free energies of the simulated DT, Ch-y, and Ch-z states, all relative to $F_{\mathrm{ST}}$ from Eq.~(\ref{Ftotal_single_twist}), as functions of $\bar{q}_0$. Simulations are run from different initial states, in order to search for the ground state. Solid lines with different colors are used to distinguish the different initial states:  red starting from DT at $\bar{q}_0=0.1$, blue starting from Ch-z at $\bar{q}_0=4.2$, cyan starting from Ch-y at $\bar{q}_0=2.8$. Arrows indicate the direction of changing $\bar{q}_0$ in each series of simulations. (b)~Positive eigenvalue of $\boldsymbol{\Delta}$ at the center of the cylinder, in the simulated ground state, as a function of $\bar{q}_0$. (c)~Free energy of the simulated ground state, relative to the free energy $F_{PDT}$ of unachievable perfect double twist, as a function of $\bar{q}_0$.}
\end{figure}

Because the $\boldsymbol{\Delta}$ deformation mode does not have rotational symmetry about the director, it can be used to characterize quantitatively how the rotational symmetry is broken as $\bar{q}_0$ increases.  Figure~2(b) shows the positive eigenvalue of $\boldsymbol{\Delta}$ at the center of the disk, in the ground state, as a function of $\bar{q}_0$. From this figure, we can see that the $\boldsymbol{\Delta}$ deformation is always zero at the center in the DT state, because that state has full rotational symmetry. When DT transforms into Ch-y, there is a discontinuous jump in the value of $\boldsymbol{\Delta}$ at the center, which implies that the full rotational symmetry is broken. If we regard $\boldsymbol{\Delta}$ at the center as an order parameter for the cholesteric phase, then the discontinuous change in $\boldsymbol{\Delta}$ also represents a first-order transition, in agreement with the free energy change in Fig.~2(a).

Overall, our director field simulations show that the local optimum structure of chiral liquid crystals, which is double twist, can fit in a finite cylinder when $\bar{q}_0=q_0 R$ is sufficiently small.  However, when $\bar{q}_0$ is large, the frustration between the local optimum structure and the spatial geometry becomes too great.  Because of this frustration, the chiral liquid crystal combines the favorable double twist with other elastic modes.  It does not form the standard cholesteric configuration everywhere in the cylinder.  Instead, it forms the Ch-y or Ch-z configuration, which have cholesteric single twist in the bulk, and more complex structure in the boundary layer near the surface. As the $\bar{q}_0$ continues to increase, the bulk grows larger relative to the boundary layer, and hence the liquid crystal asymptotically approaches the standard cholesteric configuration.

Recently, Meiri and Efrati developed a general theoretical formalism to describe cumulative geometric frustration~\cite{meiri2021cumulative}. One key feature of their theory is that the total free energy grows superextensively with respect to system size, for small system sizes. We can apply their theory to our current simulations. In their theory, the free energy is measured relative to a specific baseline, which is the free energy of the \emph{unachievable} perfect state, i.e. the ideal local free energy density times the volume. Here, this baseline is the free energy of \emph{unachievable} perfect double twist $F_{PDT}=-\pi K_{22}^2\bar{q}_0^2/[2(K_{22}-K_{24})]$, which is the free energy density $f_{DT}$ of Eq.~(\ref{fDT}) times the volume of the cylinder (per length in the $z$ direction).  Figure~2(c) shows the simulated free energy in the ground state, relative to $F_{PDT}$, as a function of $\bar{q}_0=q_0 R$, plotted on a logarithmic scale.  For small $\bar{q}_0$, the results are well fit by the prediction of Eq.~(\ref{Ftotal_double_twist}), which gives
\begin{equation}
\Delta F = F_{DT}-F_{PDT}=\frac{\pi K_{22}^4 K_{33}(q_0 R)^4 }{64(K_{22}-K_{24})^4}\to\frac{\pi\bar{q}_0^4}{4}
\end{equation}
for our simulated case of $K_{11}=K_{22}=K_{33}=2K_{24}=K=1$.  This free energy scales superextensively with system size for small size.  For large $\bar{q}_0$, the results are well fit by the prediction of Eq.~(\ref{Ftotal_single_twist}), which gives
\begin{equation}
\Delta F = F_{ST}-F_{PDT}=\frac{\pi K_{22}K_{24}(q_0 R)^2}{2(K_{22}-K_{24})}\to\frac{\pi\bar{q}_0^2}{2}.
\end{equation}
That free energy scales extensively with system size for large size.  Indeed, the transition from DT to Ch-y can be understood as a way of relieving the superextensive free energy of double twist.  Hence, our results show a crossover from superextensive to extensive growth of free energy, at a length scale set by the natural twist $q_0$, in agreement with the theory of cumulative geometric frustration.

\subsection{Nematic Order Tensor Simulations}

As an alternative possible response to geometric frustration, instead of forming a combination of director deformation modes, a liquid crystal might form regions of the favorable mode separated by disclinations~\cite{selinger2022director}.  Indeed, chiral liquid crystals often form blue phases, which can be regarded as cubic lattices of double-twist tubes and disclinations.  Here, we want to understand how this response can occur in the simple geometry of a chiral liquid crystal in a cylinder with free boundary conditions.  The formation of disclinations cannot be described by simulations of the director field.  Instead, we must use simulations of the full nematic order tensor.

We implement a series of simulations for the nematic order of a chiral liquid crystal confined in the same geometry as in the previous section, which is a cylinder of radius $R$ with free boundary conditions.  We still assume that the nematic order is a function of $x$ and $y$, independent of $z$.  It is represented by a traceless symmetric tensor $Q_{ij}(x,y)$, which has five degrees of freedom.  Wherever the liquid crystal is uniaxial, the relation between the nematic order tensor and the director is $Q_{ij} = S(\frac{3}{2}n_i n_j-\frac{1}{2}\delta_{ij})$, where $S$ is the scalar order parameter. In terms of the nematic order tensor, the Landau-de Gennes free energy of the liquid crystal (per length in the $z$ direction) is conventionally written as
\begin{align}\label{ftotal_Q_tensor}
  \nonumber F&=\int\bigg[ \frac{A_0}{2}\left(1-\frac{\gamma}{3}\right)\Tr\boldsymbol{Q}^2
  - \frac{A_0\gamma}{3}\Tr\boldsymbol{Q}^3
  + \frac{A_0\gamma}{4}\left(\Tr\boldsymbol{Q}^2\right)^2\\
  &\qquad + \frac{L_1}{2}\left(\partial_i Q_{jk}\right)\left(\partial_i Q_{jk}\right)
  + \frac{L_4}{2}\epsilon_{lik}Q_{lj}\partial_k Q_{ij}\bigg] dx dy
\end{align}
Inside the integrand, the first three terms give the bulk free energy density in terms of the parameters $A_0$ and $\gamma$. When the liquid crystal is uniform, the optimum scalar order parameter $S_0$ can be expressed as a function of $\gamma$. The last two terms in Eq.~(\ref{ftotal_Q_tensor}) give the distortion free energy density associated with spatial variations of the nematic order. The $L_1$ term is a nonchiral energy penalty for gradients of $Q_{ij}$, which corresponds to the single elastic constant $K$ for gradients of $\hat{\boldsymbol{n}}$. The $L_4$ term is a chiral term, which favors twist of the nematic order. Based on the combination of these two terms, the natural twist of a cholesteric helix is $q_0 = L_4/(4L_1)$.

Previous studies have already considered the free energy of Eq.~(\ref{ftotal_Q_tensor}) as a model for blue phases in an \emph{infinite} system~\cite{grebel1983landau,dupuis2005numerical,alexander2006stabilizing,duzgun2018comparing}. By rescaling the length, energy, and $Q$-tensor, those studies found that the behavior depends only on two dimensionless ratios
\begin{eqnarray}\label{two_dimensionless_ratios}
  \tau = \frac{9\left(3-\gamma\right)}{\gamma},\quad 
  \kappa = \sqrt{\frac{27 L_4^2}{4 L_1 A_0 \gamma}},
\end{eqnarray}
known as the reduced temperature and the chirality, respectively. Here, we want to find the behavior in a \emph{finite} cylinder of radius $R$. In addition to $\tau$ and $\kappa$, the behavior must also depend on a third dimensionless parameter related to the finite size $R$.  This third parameter can be written as $\bar{q}_0 = q_0 R = L_4 R/(4L_1)$, just as in director simulations of the previous section. In our nematic order tensor simulations, we fix the radius $R=10$ and elastic constant $L_1=2.32$, and we tune the chiral coefficient $L_4$ to explore the structural evolution. To change the free energy cost of disclinations, we adjust the values of $A_0$ and $\gamma$.  This choice of parameters implicitly sets $\tau$ and $\kappa$, and hence puts us in the cholesteric region or the blue-phase region of the known phase diagram for the infinite system.

To model pure relaxational dynamics, we define the dissipation function
\begin{eqnarray}\label{dissipation_function_for_Qtensor}
 D = \frac{1}{2}\Gamma_1 \int dx dy \dot{Q}_{ij} \dot{Q}_{ij},
\end{eqnarray}
where $\Gamma_1$ is analogous to a rotational viscosity for the $\boldsymbol{Q}$-tensor, choosing time units so that $\Gamma_1=1$.  We then solve the overdamped equations of motion
\begin{eqnarray}\label{equation_of_motion_Qtensor}
 -\frac{\delta F}{\delta Q_{ij}} - \frac{\delta D}{\delta\dot{Q}_{ij}} = 0
\end{eqnarray}
by finite-element modeling.  To avoid trapping the $\boldsymbol{Q}$-tensor field in a metastable state, we begin our simulations with different initial states, and compare the total free energies of the resulting equilibrium states to identify the overall ground state. Because introducing a disclination might require overcoming a high energy barrier, we consider an initial state with lattice of disclinations, as in the literature on blue phases~\cite{grebel1983landau},
\begin{align}\label{initial_condition_singledisclination}
\boldsymbol{Q}&=
c\left[\begin{matrix}
-1 & 0 & 0 \\
0 & -1 & 0 \\
0 & 0 & 2
\end{matrix}\right]
+b\text{Re}\left[\begin{matrix}
0 & 0 & 0 \\
0 & 1 & -i \\
0 & -i & -1
\end{matrix}\right]e^{i(qx-\frac{\pi}{3})}\nonumber\\
&\quad+b\text{Re}\left[\begin{matrix}
\frac{3}{4} & \frac{\sqrt{3}}{4} & -\frac{i\sqrt{3}}{2} \\
\frac{\sqrt{3}}{4} & \frac{1}{4} & -\frac{i}{2} \\
-\frac{i\sqrt{3}}{2} & -\frac{i}{2} & -1
\end{matrix}\right]e^{i(\frac{qx}{2}-\frac{qy\sqrt{3}}{2}+\frac{\pi}{3})}\nonumber\\
&\quad+b\text{Re}\left[\begin{matrix}
\frac{3}{4} & -\frac{\sqrt{3}}{4} & \frac{i\sqrt{3}}{2} \\
-\frac{\sqrt{3}}{4} & \frac{1}{4} & -\frac{i}{2} \\
\frac{i\sqrt{3}}{2} & -\frac{i}{2} & -1
\end{matrix}\right]e^{i(\frac{qx}{2}+\frac{qy\sqrt{3}}{2}+\frac{\pi}{3})}.
\end{align}
In this initial state, the scalar order parameter and the biaxiality are controlled by the two parameters, $b$ and $c$, and the twist deformation is tuned by $q$.

\begin{figure}
\centering
\includegraphics[width=0.48\textwidth]{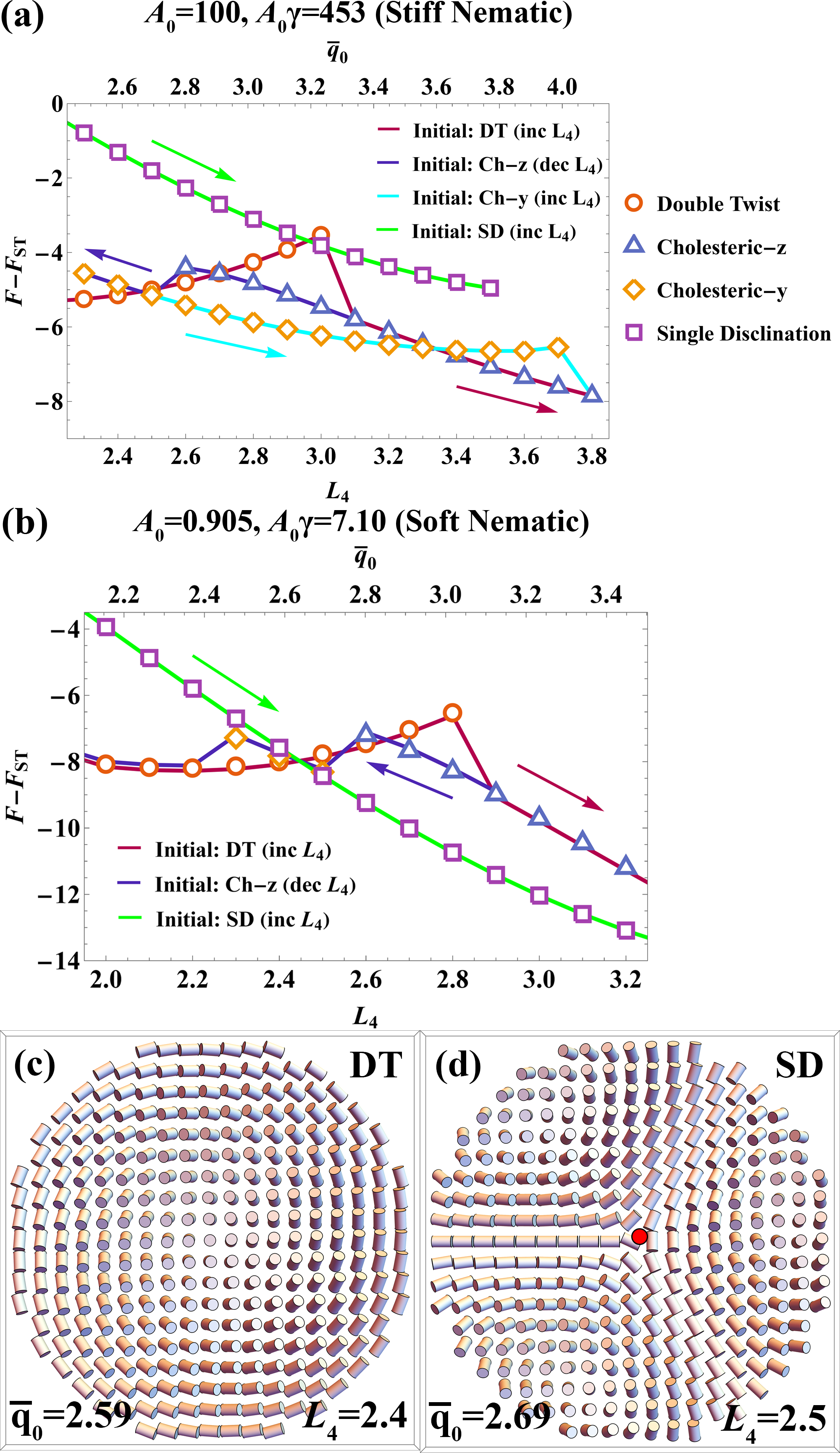}
\caption{(a)--(b)~Free energies of the states found in our nematic order tensor simulations, as functions of the chiral parameter $L_4$, with bulk coefficients for stiff and soft nematic order, respectively. All free energies are relative to the free energy of the standard biaxial cholesteric helical structure inside the cylinder, at the same value of $L_4$. As in Fig.~2, solid lines with different colors represent simulations run from different initial states. Arrows indicate the direction of changing $L_4$ in each series of simulations. (c)--(d)~Director field visualizations for the ground state with soft nematic order at $L_4=2.4$ and $2.5$, respectively. The orientation of the cylinders represents the local eigenvector of $\boldsymbol{Q}$ with the largest eigenvalue. The red dot marks the $-1/2$ disclination.}
\end{figure}

To simulate a chiral liquid crystal with stiff nematic order, we choose the bulk coefficients $A_0=100$ and $A_0\gamma=453$. When $L_4$ is changed continuously from $2$ to $3.8$, the dimensionless ratio $\tau$ remains at $-3$, and $\kappa$ ranges from $0.16$ to $0.3$.  These values of $\tau$ and $\kappa$ correspond to the cholesteric region of the phase diagram for an infinite system~\cite{alexander2006stabilizing}. Our simulation results for this range of parameters are shown in Fig.~3(a). From the free energy graph, we can identify the same structural transitions as already seen in the director field simulations. When $\bar{q}_0$ is small, the ground state is DT. As $\bar{q}_0$ is increased to about $2.7$, the ground state changes from DT to Ch-y. When $\bar{q}_0$ reaches $3.7$, there is a further transition from Ch-y to Ch-z. When $\bar{q}_0$ is large enough, the ground state becomes similar to a standard cholesteric helical structure.

To simulate a chiral liquid crystal with softer nematic order, we choose $A_0=0.905$ and $A_0\gamma=7.10$. When $L_4$ is changed continuously from $2$ to $3.2$, $\tau$ remains at $-5.6$, and $\kappa$ ranges from $1.3$ to $2$.  These values of $\tau$ and $\kappa$ are in the blue-phase region of the bulk phase diagram. The simulation results for those parameters are shown in Fig.~3(b). When $\bar{q}_0$ is  small, the ground state is still DT. Unlike the results for stiff nematic order, the ground state is never Ch-y or Ch-z. Instead, there is a transition between DT (Fig.~3(c)) and a single-disclination (SD) configuration (Fig.~3(d)) as $\bar{q}_0$ is increased to about $2.7$. By visualizing the director (defined as the eigenvector of $\boldsymbol{Q}$ with the largest eigenvalue) inside the cylinder, we can see that the SD configuration is composed of three double-twist regions separated by a $-1/2$ disclination at the center. As we further increase $\bar{q}_0$ to values greater than $4.8$, the ground state becomes a complex combination of disclinations, double-twist regions, and single-twist regions.

To explore the extreme case in which the free energy cost of a disclination is very low, we perform simulations with $A_0=-0.027$ and $A_0\gamma=2.92$. Figure~4 shows the ground states for three values of $\bar{q}_0$ beyond the value where a single disclination forms. At $\bar{q}_0=4.31$, the ground state (Fig.~4(a)) shows seven large double-twist regions separated by six $-1/2$ disclinations. As $\bar{q}_0$ is increased to $6.47$, there is a complex arrangement of double-twist structures inside the cylinder. The ground state (Fig.~4(b)) has three large double-twist regions separated by a $-1/2$ disclination at the center, which resembles an SD configuration except that six small double-twist regions are inserted close to the surface. When $\bar{q}_0=8.62$, the size of the optimum double-twist region becomes small enough, compared to the cylinder radius, so that the ground state (Fig.~4(c)) is filled with many double-twist domains separated by $-1/2$ disclinations. The general behavior here is that, as $\bar{q}_0$ increases, the chiral liquid crystal tries to fill the cylinder with as many double-twist domains as possible.

\begin{figure*}
\centering
\includegraphics[width=0.75\textwidth]{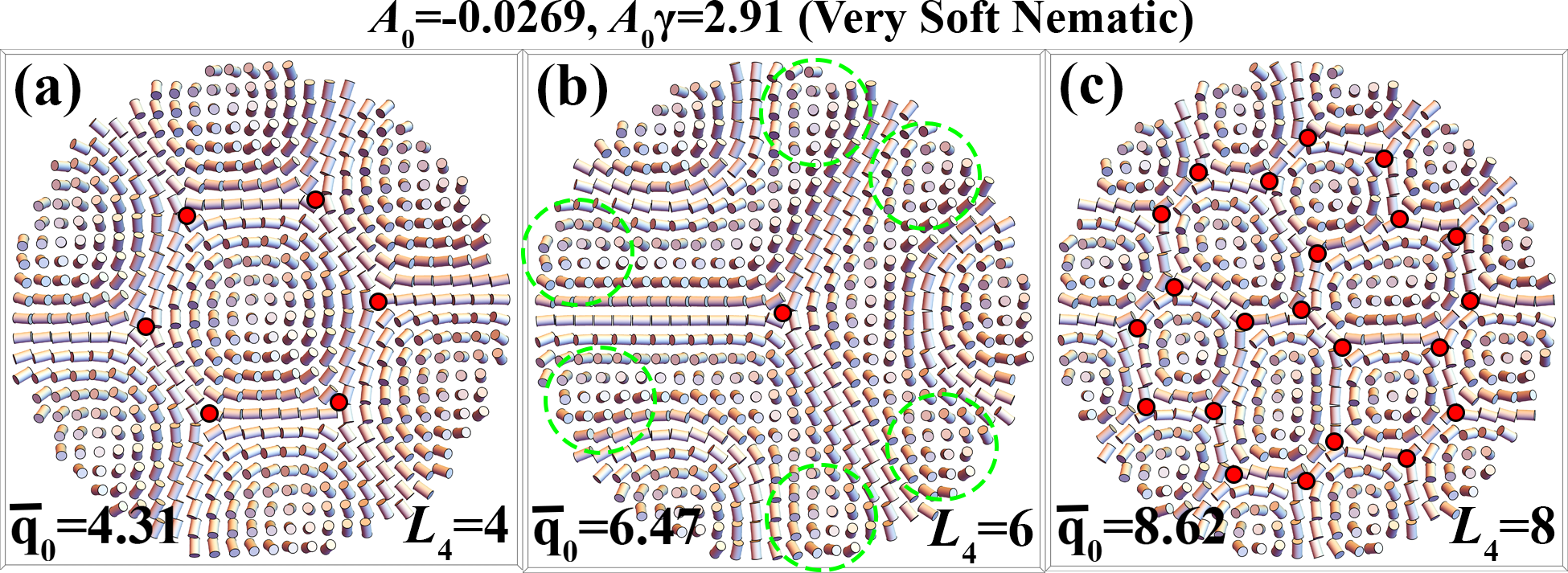}
\caption{(a)--(c) Visualization of the nematic order tensor fields for a chiral liquid crystal with very soft nematic order at $L_4=4$, $6$, and $8$, respectively. The orientation of the cylinders represents the local eigenvector of $\boldsymbol{Q}$ with the largest eigenvalue. The red dots mark the $-1/2$ disclinations, and the green circles indicate the small double-twist regions in part~(b).}
\end{figure*}

Overall, our simulations of the nematic order tensor demonstrate that a chiral liquid crystal can have two possible responses to geometric frustration.  If the bulk free energy coefficients are large, so that the nematic order is stiff, then the liquid crystal combines the favorable double twist with the unfavorable $\boldsymbol{\Delta}$ mode, and it forms a cholesteric structure with single twist.  It does not form disclinations, because the energy cost of disclinations is too high.  These results are consistent with our director simulations in the previous section.  By contrast, if the bulk free energy coefficients are smaller, so that the nematic order is softer, then the liquid crystal forms regions of the favorable double twist separated by disclinations.  In that sense, it forms a simple version of a blue phase in the finite geometry.

\section{Chiral Liquid Crystal in a Slab}

In this section, we consider a slab of chiral liquid crystal between two infinite, parallel plates, with free boundary conditions on both of the plates. If there were no geometric frustration, then we might expect the liquid crystal to form a standard cholesteric director field. Here, we show that geometric frustration leads to a more complex structure near the plates.

To model the chiral liquid crystal between the plates, we use a Cartesian coordinate system with the plates parallel to the $(x,y)$ plane, located at $z=\pm d/2$, respectively. To simplify the model, we assume that the director field depends only on $x$ and $z$, and is independent of $y$. We write it in conventional spherical coordinates as 
\begin{eqnarray}\label{director_field_n_in_a_slab}
  \hat{\boldsymbol{n}}(x,z) = \left(\sin{\theta}\cos{\phi}, \sin{\theta}\sin{\phi}, \cos{\theta}\right).
\end{eqnarray}
The free energy takes the same form as Eq.~(\ref{ftotal_director_field_simulations}), and the dissipation function takes the same form as Eq.~(\ref{dissipation_function_for_director_field}), except now expressed in terms of angles $\theta(x,z)$ and $\phi(x,z)$. We solve the overdamped equations of motion for $\theta$ and $\phi$ to obtain the equilibrium director field.

\begin{figure*}
\centering
\includegraphics[width=0.9\textwidth]{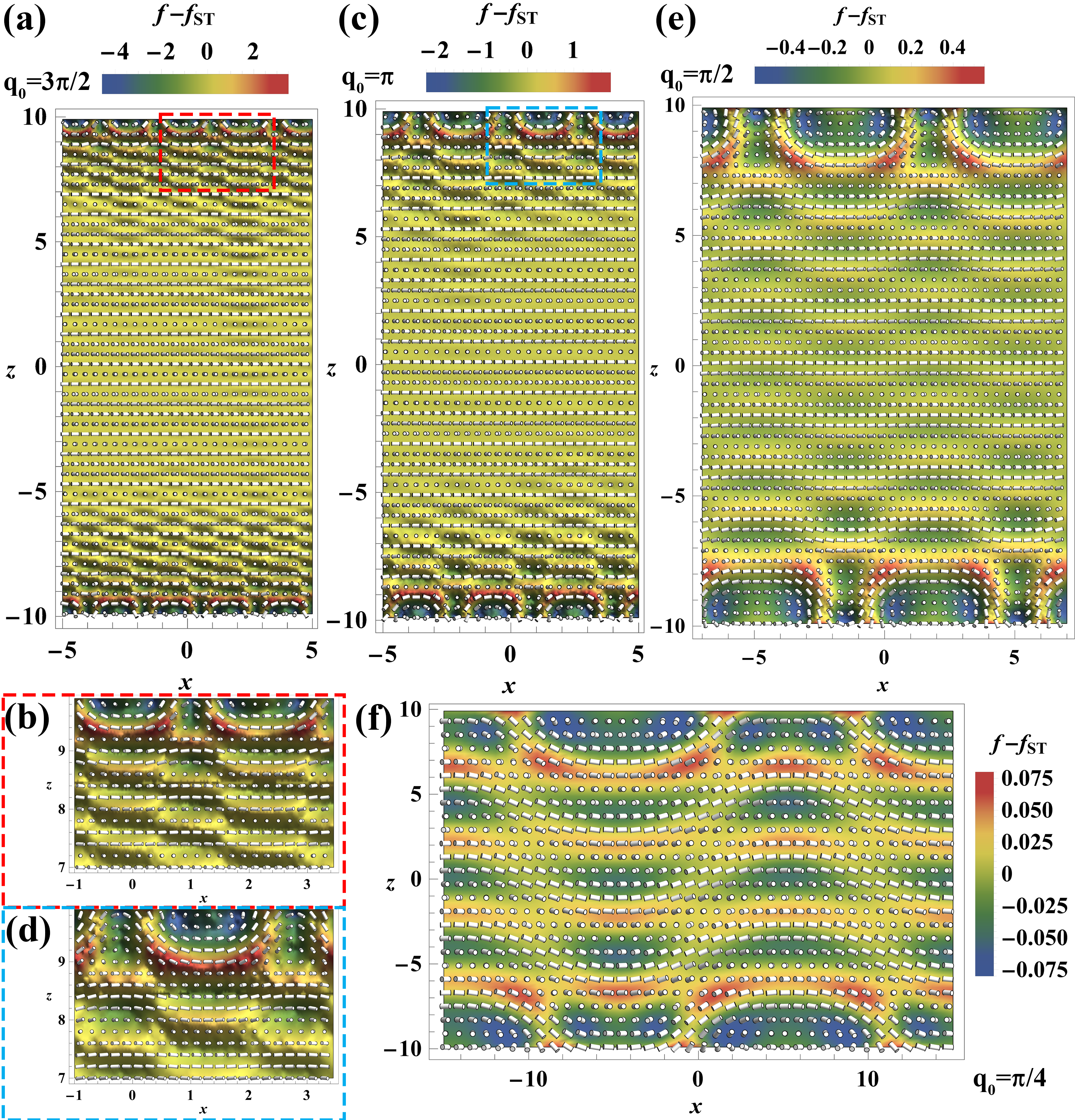}
\caption{\label{fig:5} Visualization of the director field in chiral liquid crystals confined between parallel plates with free boundary conditions. In each image, the orientation of the cylinders indicates the local director $\hat{\boldsymbol{n}}$, and the color below the cylinders shows the free energy density (relative to the standard cholesteric helix at the corresponding value of $q_0$). (a)~Full height of cell with $q_0=3\pi/2$. (b)~Zoomed-in surface state for $q_0=3\pi/2$. (c)~Full height of cell with $q_0=\pi$. (d)~Zoomed-in surface state for $q_0=\pi$. (e)~Full height of cell with $q_0=\pi/2$. (f)~Full height of cell with $q_0=\pi/4$. Compared to the previous director field simulations in a capillary tube, we use twist with the opposite handedness in these simulations.}
\end{figure*}

Simulations are conducted in a rectangular box in the $(x,z)$ plane. In the $x$ direction, the size of the box is $40$, and periodic boundary conditions are applied to simulate an unbounded system. In the $z$ direction, the thickness of the box is fixed to $d=20$, and we vary $q_0$ to study how the director configuration depends on the ratio of the cell thickness to the natural pitch. For the initial condition, we choose a standard cholesteric helical structure with a pitch slightly bigger than the natural pitch $\pi/q_0$, with pitch axis parallel to $\hat{\boldsymbol{z}}$. By relaxing the director field from the initial condition, we find equilibrium configurations which have lower total free energy than the standard cholesteric helical structure at the corresponding $q_0$.  Our results are illustrated in Fig.~5.

Let us first consider the case of $q_0 = 3\pi/2$ (Fig.~5(a)).  In the interior of the cell, the director field shows a standard cholesteric helical structure, with pitch axis parallel to $\hat{\boldsymbol{z}}$.  The free energy density in the interior is equal to the free energy density of the standard cholesteric phase at $q_0=3\pi/2$. By contrast, at each of the free boundaries, there is a row of large semicircular domains of double twist, arranged regularly in the $x$ direction, forming a surface state. Between each pair of two adjacent semicircular domains (Fig.~5(b)), a small double-twist region is inserted to connect the entire chain together. At the center of each large semicircular domain, the free energy density is lower than the standard cholesteric phase. Away from the center, the free energy density gradually increases. Approaching the domain periphery, the free energy density becomes even higher than the standard cholesteric phase. However, the total free energy for each domain is still lower than the total free energy of the standard cholesteric phase filling the same region. To connect the domains near the free surfaces and the cholesteric helical structure in the interior, the director field transforms gradually from the surface state to the interior structure, manifested as a trail of wrinkles in the free energy density plot.

For $q_0 = \pi$ (Fig.~5(c)), the equilibrium state is similar to the previous case, except that the semicircular domains of double twist (Fig.~5(d)) become larger. Because of the increased size of these domains, the surface state has a longer wavelength in the $x$ direction. As $q_0$ is decreased to $\pi/2$ (Fig.~5(e)), the semicircular domains of double twist become even larger, so that the surface states fully penetrate through the interior of the cell. The director field around $z=0$ is slightly distorted away from the standard cholesteric helical structure because of the influence of the surface states. When $q_0 = \pi/4$ (Fig.~5(f)), the surface states strongly interfere with each other, resulting in a buckled lamellar pattern in the interior of the liquid crystal cell.

In a sense, the behavior found here can be regarded as an inside-out version of the Helfrich-Hurault effect.  In the usual Helfrich-Hurault effect, which is observed in many cholesteric liquid crystals~\cite{blanc2021helfrich}, the surfaces provide rigid anchoring.  A horizontal modulation forms in the interior of a cell, while the surfaces maintain a standard cholesteric helical structure.  Here, the surfaces have free boundary conditions, and a horizontal modulation forms at the surfaces, while the interior remains closer to the standard cholesteric helix.

\section{Discussion}

In this work, we theoretically demonstrate that double twist is the optimum deformation that minimizes the local free energy density of chiral liquid crystals. Because of geometric constraints, pure double twist cannot fill up 3D Euclidean space, and chiral liquid crystals must compromise between the local optimum and the global structure. We have studied two model systems to see how this geometric frustration affects chiral liquid crystals confined in a finite system with free boundary conditions.

First, we investigate a chiral liquid crystal confined in a long cylinder, using analytic theory, director field simulations, and nematic order tensor simulations. All these techniques show that the equilibrium structure is controlled by the dimensionless parameter $\bar{q}_0=q_0 R$, i.e.\ the ratio of the cylinder radius to the natural pitch. When $\bar{q}_0$ is sufficiently small, the liquid crystal forms a double-twist configuration. However, when $\bar{q}_0$ is larger, the double-twist configuration accumulates too much geometric frustration.  In that case, if the nematic order is stiff, the liquid crystal forms a cholesteric helix, which combines the favorable double twist with the unfavorable $\boldsymbol{\Delta}$ deformation mode. If the nematic order is soft, it forms double-twist domains separated by disclinations.

Second, we study a chiral liquid crystal confined between two infinite, parallel plates with free boundary conditions. In this geometry, double twist cannot fill up the slab. Instead, geometric frustration induces surface states close to the free boundaries, where semicircular domains of double twist are arranged in a row. The size of each semicircular domain at the surface becomes larger as the natural pitch increases. When the ratio of the cell thickness to the natural pitch becomes small enough, the surface states penetrate through the entire cell, causing a buckled lamellar structure in the interior.

After this research was completed, we learned of a recent experimental and theoretical article by Pi\v{s}ljar et al.~\cite{pisljar2022skyrmions}, which investigates blue phase III as a topological fluid of skyrmions. Their work is related to our current study because it also examines finite-size effects in chiral liquid crystals. When they reduce the thickness of a sample, they find that the temperature range of the cholesteric phase is reduced, and the liquid crystal can more easily form half-skyrmions or a blue phase. This is the same trend that we find in this article. It supports the view that the cholesteric phase is a frustrated structure, which is only stabilized because of geometric constraints in a large geometry.

To compare our work with Ref.~\cite{pisljar2022skyrmions}, we would highlight two small distinctions, which are only differences of emphasis, not scientific disagreements. First, they present a detailed study of a realistic experimental system.  By contrast, we investigate very simple models, with the minimal features to demonstrate geometric frustration. Second, by concentrating on skyrmions or half-skyrmions, they emphasize the effects of topology. We would say that the fundamental issue is the elasticity of chiral liquid crystals, which favors double twist more than cholesteric single twist, combined with geometric constraints. Topology is important only because disclinations allow the liquid crystal to develop more regions of double twist.

In conclusion, by studying two model systems with free boundary conditions, we explicitly demonstrate geometric frustration in chiral liquid crystals. Because of frustration, the configuration of a chiral liquid crystal in a finite system can depend on the geometry of the container in unexpected ways. This theoretical research provides a new perspective on how to understand the structure of chiral liquid crystals, and may help to predict and control new geometric effects.

\section*{Conflicts of interest}

There are no conflicts to declare.

\section*{Acknowledgements}

This work was supported by National Science Foundation Grant DMR-1409658. 

%%%END OF MAIN TEXT%%%

%The \balance command can be used to balance the columns on the final page if desired. It should be placed anywhere within the first column of the last page.

%\balance

%If notes are included in your references you can change the title from 'References' to 'Notes and references' using the following command:
%\renewcommand\refname{Notes and references}

%%%REFERENCES%%%
\bibliography{version3} %You need to replace "rsc" on this line with the name of your .bib file
\bibliographystyle{rsc} %the RSC's .bst file

\end{document}